# Computational design of a new layered superconductor LaOTlF$_2$

Zhihong Yuan, Jingjing Meng, Rui Liu, Pengyu Zheng, Xiaobo Ma, Guangwei Wang, Tianye Yu, Yiran Peng, and Zhiping Yin*

*Department of Physics and Center for Advanced Quantum Studies, Beijing Normal University, Beijing 100875, China.*

A new layered compound LaOTlF$_2$ is designed and investigated using first-principles calculations in this work. The parent compound is an insulator with an indirect band gap of 2.65 eV. Electron-doping of the parent compound makes the material metallic. In the meantime, several lattice vibrational modes couple strongly to the conduction band, leading to a large electron-phonon coupling constant and conventional superconductivity. The highest superconducting transition temperature $T_c$ is predicted to be approximately 8.6 K with λ about 1.25 in the optimally doped LaO$_{0.95}$F$_{0.05}$TlF$_2$, where λ is calculated using the Wannier interpolation technique.

## 1 Introduction

Since the discovery of superconductivity in Hg in 1911,[1] searching for new high-temperature superconductors and studying the superconducting mechanism have always been hot topics in the community of condensed matter physics. Layered materials have generated substantial interest in superconducting research due to their abundant structural variations, showing that superconductivity can be induced and improved through doping, applying pressure, intercalation, *etc.*.[2-6]

High-$T_c$ cuprates[4] and Fe-based[5] compounds have triggered numerous studies focusing on elucidating the mechanism of superconductivity and increasing their superconducting transition temperature $T_c$. A noteworthy feature of these compounds is their layered crystal structure consisting of alternately stacked functional layers (such as CuO$_2$ or FeAs layers) and various filling layers. Motivated by the structure of LaOFeAs, the first BiS$_2$-based layered superconductor Bi$_4$O$_4$S$_3$ was reported in 2012,[7] and analogous LaOBiS$_2$ was immediately studied, which possesses a layered structure composed of BiS$_2$ layers and LaO layers.[8] The parent phase of LaOBiS$_2$ is an insulator, whereas partially substituting O with F induces superconductivity, and the optimally doped LaO$_{0.5}$F$_{0.5}$BiS$_2$ shows superconductivity at a $T_c$ = 2.7 K under ambient pressure.[8] To date, a series of related compounds have been synthesized by altering the filling layer and doping electrons through the use of different elements, while enormous efforts have been dedicated to exploring and improving their superconductivity.[9-12]

As mentioned above, layered materials have highly contributed to the rapid development of superconductivity, but the superconducting mechanism of most layered compounds is still unclear; therefore, it is very important to design new layered superconducting materials and study their properties, which may be an opportunity for people to unravel the mystery of superconductivity. Currently, material design through computational simulation is an important way to search for new materials.[13-16]

In 2013, a class of BaBiO$_3$-like thallium halide-based compounds ATlX$_3$ (A = Cs, Rb, K; X = F, Cl) was rationally designed by Yin and Kotliar[13] as candidates for new high-temperature superconductors. Following their prediction, CsTlF$_3$ and CsTlCl$_3$ were successfully synthesized in laboratory by Retuerto and collaborators.[17] However, hole-doping these bulk materials has been unsuccessful. To overcome the difficulties of doping in bulk materials, in this work, we aim to design thallium halide-based layered compounds, and investigate their electronic structures, lattice dynamics and superconducting properties.

## 2 Computational methods

Two crystal structure prediction tools, the swarm intelligence-based CALYPSO (Crystal structure AnaLYsis by Particle Swarm Optimization)[18] and USPEX (Universal Structure Predictor: Evolutionary Xtallography),[19] are employed to find the ground state crystal structure of LaOTlF$_2$. To model the electron-doping system, we replace O with F through the virtual crystal approximation (VCA) method.[20,21] The weighted average mass at each doping level is taken as the atomic mass of the virtual atom.

The structural, electronic, and lattice-dynamical properties are obtained with density functional theory (DFT) and density functional perturbation theory (DFPT)[22] using the Quantum ESPRESSO code[23] with norm-conserving pseudopotentials.[24,25] The Perdew-Burke-Ernzerhof form of the generalized gradient approximation[26] is used for the exchange-correlation functional. After a full convergence test, the kinetic energy cutoff and the charge density cutoff of the plane wave basis are chosen to be 100 Ry and 400 Ry, respectively. Brillouin-zone grids with up to 30 × 30 × 10 $k$-points and a Marzari-vanderbilt[27] smearing broadening as small as 0.005 Ry are used for the Fermi surface and band structure calculations. The phonon and phonon perturbation potentials are calculated on a Γ-centered 8 × 8 × 2 $q$ mesh and 16 × 16 × 4 $k$-point grid within the framework of DFPT.[22]

The superconducting properties of $LaO_{1-x}F_xTlF_2$ are studied within the EPW code,[28,29] which uses DFPT and the maximally localized Wannier functions theory to calculate the electron-phonon couplings and related properties. We use six maximally localized Wannier functions (three p states for each Tl atom) to describe the band structure around the Fermi level. First, the electronic band energy, and electron-phonon coupling (EPC) matrix elements are calculated on coarse $k$- and $q$-meshes of 8 × 8 × 4 and 8 × 8 × 2, respectively. Then, according to the convergence test of the EPC constant, 144 × 144 × 16 and 24 × 24 × 8 fine electron and phonon grids are used to interpolate the EPC properties with EPW codes,[28,29] and the Dirac δ-functions for electrons and phonons are replaced by smearing functions with widths of 10 meV and 0.2 meV, respectively.

## 3 Results and discussion

### 3.1 Structural properties

We found no information about the $LaOTlF_2$ compound in the Inorganic Crystal Structure Database (ICSD);[30] therefore, it is a new material to our best knowledge. We used two crystal structure prediction tools, namely, CALYPSO[18] and USPEX[19] to search the crystal structure of $LaOTlF_2$. Both tools predicted that the crystal structure of $LaOTlF_2$ with the lowest total energy at ambient pressure is a layered structure consisting of alternately stacked LaO layers and $TlF_2$ layers, forming a tetragonal structure with $P4/nmm$ space group (No. 129), and lattice constants a = 4.171 Å and c = 11.974 Å. Fig. 1 shows the crystal structure of this compound. Each primitive unit cell contains two formula units, the La and Tl atoms are located at the 2$c$ position, O atoms are located at the 2$b$ position, while F atoms take the 4$f$ site; moreover, the Tl and F atoms lie in two separate planes. The crystal structure of $LaOTlF_2$ is drawn with the VESTA code.[31]

Generally, the stability of a hypothetical compound should be tested under the following two conditions: a) whether the hypothetical compound is energetically favorable relative to the reactants;[32] and b) whether there is unstable phonon mode in the phonon spectra. To check the energetic stability, we consider here a possible reaction to synthesize $LaOTlF_2$ as an example: LaOF + TlF ⟶ $LaOTlF_2$ + Q. By using the structures of the related reactants taken from the ICSD,[30] our calculations suggest that the reaction releases approximately 0.115 Ry energy per $LaOTlF_2$ unit (152 kJ/mol) and confirm that the above reaction is energetically favored. The dynamic stability of $LaOTlF_2$ is verified by the density functional perturbation theory (DFPT) method. The calculated phonon dispersion of the parent compound $LaOTlF_2$ is plotted in Fig. 2. The absence of imaginary frequency modes indicates the dynamic stability of $LaOTlF_2$. It is concluded that our predicted structure of $LaOTlF_2$ is energetically and dynamically stable. Therefore, from a theoretical point of view, it is highly likely that $LaOTlF_2$ can be successfully synthesized in future experiments.

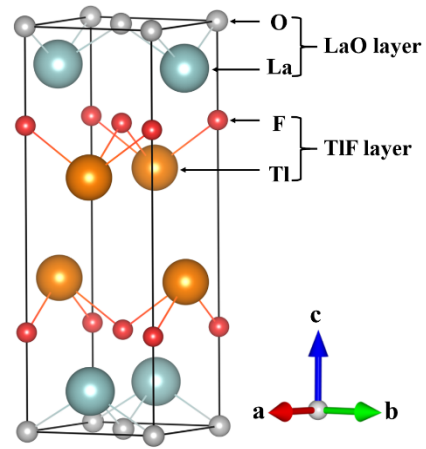

Fig. 1 Schematic image of the crystal structure of $LaOTlF_2$.

The total and partial phonon density of states depicted in Fig. 2 show that the high-frequency modes over 305 cm$^{-1}$ are contributed from the vibrations of the O atoms, while the low-frequency modes below 75 cm$^{-1}$ mostly correspond to the vibrations of the Tl atoms. The intermediate-frequency region between 75 cm$^{-1}$ and 240 cm$^{-1}$ involves the vibrations of the La and F atoms, and the frequencies beyond 240 cm$^{-1}$ up to 305 cm$^{-1}$ only relate to the vibrations of F atoms.

### 3.2 Electronic properties

Figure 3 presents the calculated electronic band structure and the density of states of $LaOTlF_2$. The highest valence band is separated from the lowest conduction band by a 2.65 eV indirect gap, indicating the insulating behavior of $LaOTlF_2$.

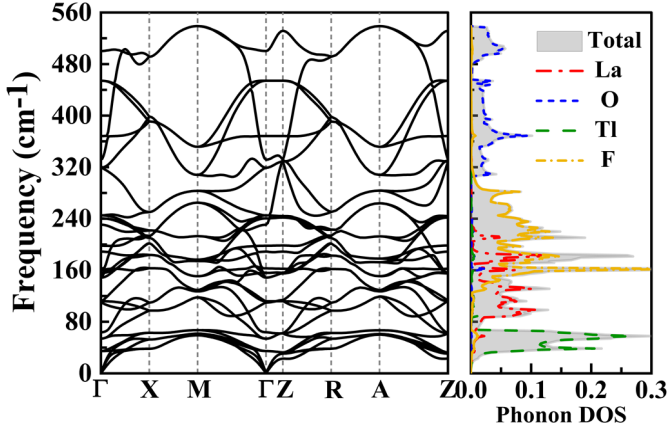

Fig. 2 Phonon spectra and phonon density of states of LaOTlF$_2$.

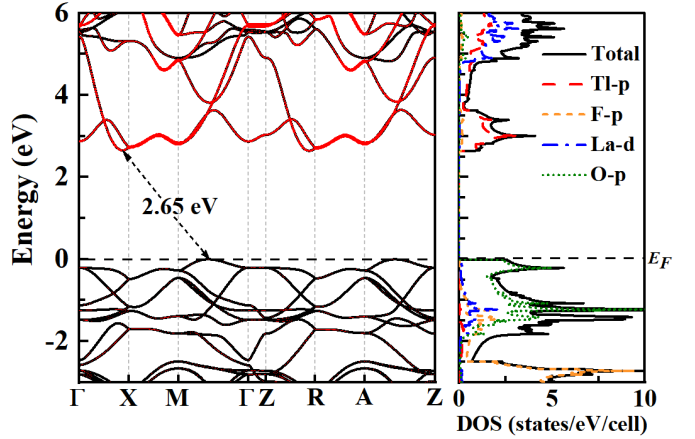

Fig. 3 Band structure and density of states of LaOTlF$_2$. (The red circles in the band structure are used to denote the Tl-6$p$ orbital character in different bands, the radius of the circle is proportional to the orbital weight, and the black horizontal dotted line represents the Fermi energy level).

According to the partial density of states, the lower part of the conduction bands is dominated by Tl-6$p$ orbital character and contains a small amount of F-2$p$ orbital character, while the La-5$d$ states lie 5 eV above the Fermi level. The valence bands around the Fermi level mainly come from the O-2$p$ orbitals, while the bands located between -1.5 and -2.5 eV come from the combination of the O-2$p$, La-5$d$ and F-2$p$ states. The major contributions of the F-2$p$ orbitals lie 2.5 eV below the Fermi level.

Electron-doping of materials is a common way to induce superconductivity.[5,6,8,33] In the following section, the effects of electron-doping of LaOTlF$_2$ are investigated by substituting O with F, namely, forming LaO$_{1-x}$F$_x$TlF$_2$ with F doping level $x$. In the crystal structure optimization, the cell shapes and the atomic coordinates are optimized within the same space group of the parent phase. The calculated band structures and corresponding Fermi surfaces (FSs) of LaO$_{1-x}$F$_x$TlF$_2$ at three different F doping levels ($x$ = 0.05, 0.13, 0.15) are shown in Fig. 4. Compared with the band structure of undoped LaOTlF$_2$, the general shape and dispersion of the bands remain almost unchanged, and the main effect of the partial substitution of O with F is that the bands with Tl-6$p$ orbital character cross the Fermi energy and exhibit a rigid-band shift around the Fermi energy; thus, the material becomes metallic. The FSs exhibit quasi-two-dimensional topology in all cases, reflecting the layered crystal structure of these compounds. At the F doping $x$ = 0.05, there are two separated elliptical cylindrical electron-like FSs around the X point [Fig. 4(a)]. As we further increase the F doping level, the two elliptical cylindrical electron-like FSs overlap with each other [Fig. 4(b) and 4(c)].

### 3.3 Lattice dynamic and electron-phonon coupling

In this section, the lattice dynamics and electron-phonon coupling of LaO$_{1-x}$F$_x$TlF$_2$ ($x$ = 0.05, 0.13) are analyzed with DFPT calculations and the Wannier interpolation technique, which has proven to be very reliable in predicting the $T_c$ of conventional superconductors.[34,35] Fig. 4 reveals that the Wannier interpolated band structures are in good agreement with those obtained by DFT calculations, which forms a solid foundation for our Electron-phonon Wannier (EPW) calculations.

Figure 5 displays the phonon dispersions of LaO$_{1-x}$F$_x$TlF$_2$ corresponding to several doping levels ($x$ = 0.05, 0.13, 0.15). The contributions to the electron-phonon coupling (EPC) from each phonon mode $\nu$ at each $q$ point ($\lambda_{q\nu}$) are denoted by red dots on the phonon dispersion lines (the radius of the red dot is proportional to the $\lambda_{q\nu}$). According to the vibration mode analysis, the out-of-plane transverse acoustic modes around the Γ point provide the largest contribution to the EPC for $x$ = 0.05 and 0.13, which results in strong softening of this transverse acoustic mode, and suggests lattice instability at higher F doping levels. Indeed, our calculations show that when the F doping level is more than 0.13 ($x$ > 0.13), the frequency of this transverse acoustic mode becomes imaginary around $q$ = (0.125, 0.125, 0) and $q$ = (0, 0.125, 0) [Fig. 5(c)].

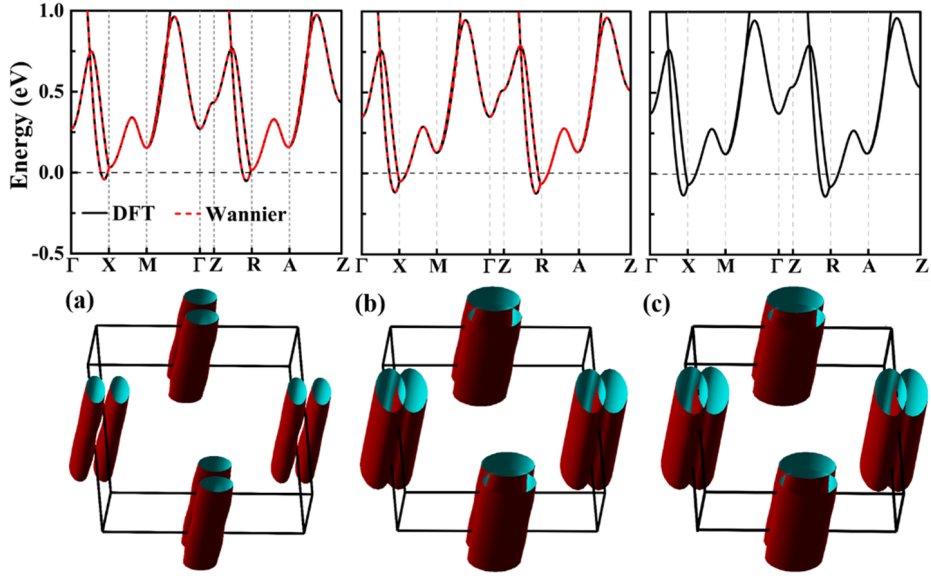

Fig. 4 Band structures (top panel) and the corresponding Fermi surfaces (bottom panel) of (a) LaO$_{0.95}$F$_{0.05}$TlF$_2$, (b) LaO$_{0.87}$F$_{0.13}$TlF$_2$ and (c) LaO$_{0.85}$F$_{0.15}$TlF$_2$. The black lines are calculated by DFT, and the red dots are obtained through Maximally localized Wannier functions (MLWF) interpolation.

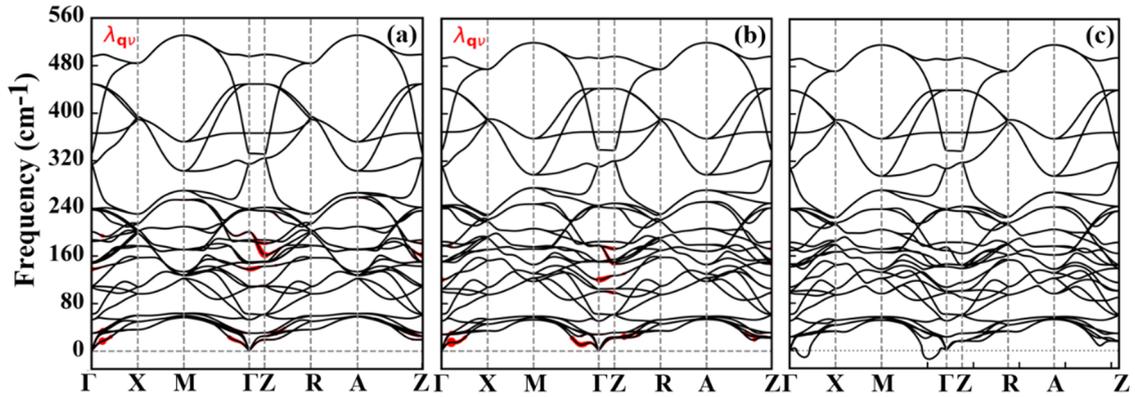

Fig. 5 Phonon spectra of (a) LaO$_{0.95}$F$_{0.05}$TlF$_2$, (b) LaO$_{0.87}$F$_{0.13}$TlF$_2$ and (c) LaO$_{0.85}$F$_{0.15}$TlF$_2$. The radius of the red dots is proportional to the corresponding mode- and momentum-dependent electron-phonon coupling $\lambda_{qv}$.

In addition, a few longitudinal optical phonon modes along the Γ - Z direction involving mostly vibrations of F atoms also have substantial EPC in the frequency range of 100 - 250 cm$^{-1}$. Taking the doping level $x = 0.13$ as an example [Fig. 5(b)], the large EPC mode around the Γ point at 120 cm$^{-1}$ corresponds to the vibrations of F atoms moving in-phase between two TlF$_2$ layers and out-of-phase within the same layer as depicted in Fig. 6(a). The large EPC mode around the Z point at 147 cm$^{-1}$ (172 cm$^{-1}$) corresponds to the vibrations of the F atoms moving in-phase (out-of-phase) between two TlF$_2$ layers and in-phase within the same layer as shown in Fig. 6(b) [Fig. 6(c)].

For further analysis, the Eliashberg function $\alpha^2F(\omega)$ and the cumulative frequency dependence of electron-phonon coupling $\lambda(\omega)$ are calculated and plotted in Fig. 7. The Eliashberg function $\alpha^2F(\omega)$ indicates that the whole frequency region can be approximately divided into three parts. Taking the doping level $x = 0.13$ as an example [Fig. 7(b)], approximately 76.2% of the total EPC is contributed by the low frequency modes (below 65 cm$^{-1}$). The $\alpha^2F(\omega)$ have sharp peaks and $\lambda(\omega)$ increases rapidly in this region. The intermediate-frequency modes (80 ~ 280 cm$^{-1}$) contribute 23.5% of the total EPC. The modes in the high frequency region (above 300 cm$^{-1}$) provide only a negligible contribution to the total EPC. By integrating $\alpha^2F(\omega)$, the total electron-phonon coupling constant λ is calculated, which is 1.25 and 1.70 for LaO$_{0.95}$F$_{0.05}$TlF$_2$ and LaO$_{0.87}$F$_{0.13}$TlF$_2$, respectively. We notice that the EPC of these two F-doped compounds are very strong (λ > 1.2); thus the $T_c$ values of these two compounds are estimated through the revised Allen-Dynes formula[36,37] with $\mu^* = 0.1$. According to our calculations, the $T_c$ values of LaO$_{0.95}$F$_{0.05}$TlF$_2$ and LaO$_{0.87}$F$_{0.13}$TlF$_2$ are 8.6 K and 7.5 K, respectively.

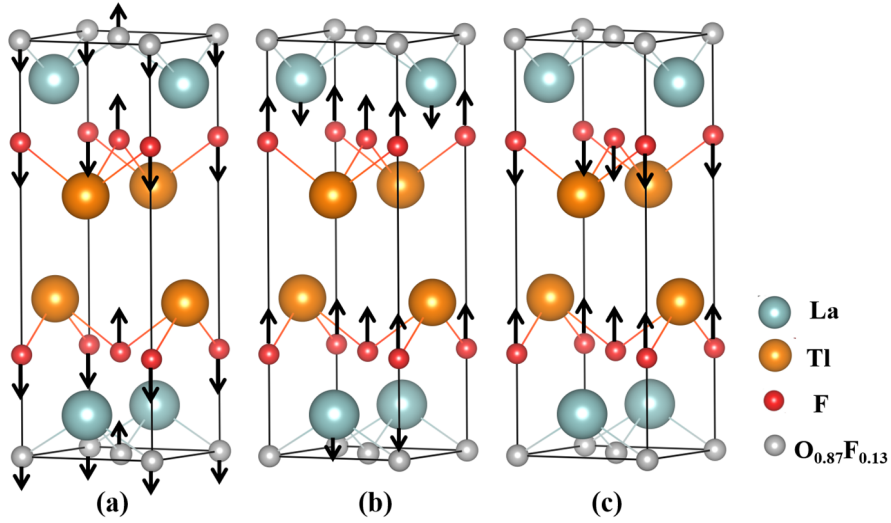

Fig. 6 Three important vibrational modes with large electron-phonon coupling: (a) The mode at Γ with phonon frequency ω~120 cm$^{-1}$. (b, c) The modes at Z with ω~147 cm$^{-1}$ and 172 cm$^{-1}$, respectively.

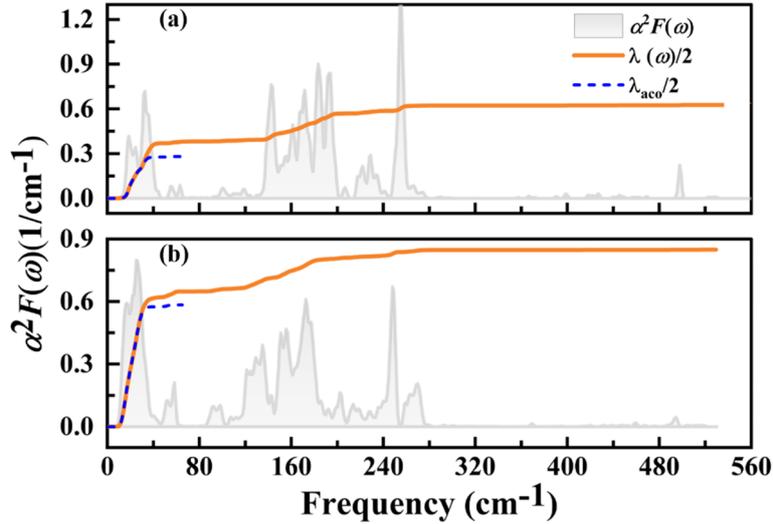

Fig. 7 The Eliashberg spectral function $\alpha^2F(\omega)$ and cumulative frequency dependence of the EPC λ(ω) of (a) LaO$_{0.95}$F$_{0.05}$TlF$_2$ and (b) LaO$_{0.87}$F$_{0.13}$TlF$_2$. The results of $\alpha^2F(\omega)$ are integrated on a homogeneous fine *k*-point grid containing 144 × 144 × 16 points, and a fine homogeneous *q*-mesh with 24 × 24 × 8 points.

## 4 Summary and outlook

In summary, a new layered compound LaOTlF$_2$ is predicted by means of first-principles calculations. It is shown to be energetically and dynamically stable in a tetragonal structure with *P*4/*nmm* space group. Electronic structure calculations suggest that the parent compound LaOTlF$_2$ is an insulator with an indirect band gap of 2.65 eV, while substitution of O with F makes the material metallic, where the metallicity is mostly contributed by Tl-6*p* electrons. Lattice dynamics and EPC calculations show that the out-of-plane transverse acoustic modes and a few longitudinal optical phonon modes along the Γ - Z direction couple strongly to the conduction band leading to a large EPC ($\lambda > 1.2$) and conventional superconductivity. The superconducting transition temperature $T_c$ of LaO$_{0.95}$F$_{0.05}$TlF$_2$ is approximately 8.6 K with $\lambda$ about 1.25, where $\lambda$ is obtained using the Wannier interpolation technique.

Owing to the layered structure of LaOTlF$_2$, by replacing its superconducting layer (TlF$_2$), we may obtain other superconductors. We tried to replace the TlF$_2$ layer with a TlCl$_2$ layer, and also explored possible superconductivity of LaOTlCl$_2$. According to our results, LaOTlCl$_2$ shares a similar crystal structure as LaOTlF$_2$: it also crystallizes in the *P*4/*nmm* space group and has lattice constants $a$ = 4.231 Å and c = 14.193 Å. The parent compound is also an insulator with an indirect band gap of 3.47 eV, but exhibits metallic behavior after doping F to O sites. The lattice dynamics and EPC calculations show that the out-of-plane transverse acoustic modes and some optical phonon modes contributed from the Tl and Cl atoms couple strongly to the conduction band leading to a large electron-phonon coupling constant and conventional superconductivity. The highest critical superconducting temperature $T_c$ is predicted to be approximately 8.1 K when doped with 15% F, and the corresponding $\lambda$ is 1.71.

In addition, by replacing the LaO layer with different layers, such as the REO layer (RE: Ce, Pr, Nd, and Sm), or partially substituting the RE atom with the Sr or Eu atom, a number of TlF$_2$ and TlCl$_2$ based compounds could be obtained. Meanwhile, various avenues for doping should also be explored, such as electrostatic gating, which has successfully induced superconductivity in some materials.[33,38] Furthermore, chemical or external pressure always plays an important role in the superconductivity of layered materials; thus, applying suitable pressure on LaO$_{1-x}$F$_x$TlF$_2$/LaO$_{1-x}$F$_x$TlCl$_2$ may enhances its superconductivity. We expect that these two new families of TlF$_2$ and TlCl$_2$ based compounds will open a new field of layered superconductors.

## Author contributions

Z.P. Yin conceived and directed this project. The crystal structures of the two compounds are predicted by Z.P. Yin and J.J. Meng. Z.H. Yuan performed the DFT, DFPT and EPW calculations. Z.H. Yuan and Z.P. Yin wrote the manuscript with input from R. Liu and P.Y. Zheng. X.B. Ma, G.W. Wang, T.Y. Yu and Y.R. Peng contributed to analyzing the data and discussing the results. All authors commented on the manuscript.

## Conflicts of interest

There are no conflicts to declare.

## Acknowledgements

This work was supported by the National Natural Science Foundation of China (Grant No. 12074041 and 11674030), the Fundamental Research Funds for the Central Universities (Grant No.310421113), the National Key Research and Development Program of China through Contract No. 2016YFA0302300, and the start-up funding of Beijing Normal University. The calculations were carried out with high performance computing cluster of Beijing Normal University in Zhuhai.

*Requests for materials should be addressed to Z.P.Y. at yinzhiping@bnu.edu.cn

## Notes and references